\documentclass[aps,prl,reprint,superscriptaddress]{revtex4-1}
\usepackage{graphicx}
\usepackage{amsmath}
\usepackage{amssymb}

\begin{document}
\title{Vortex distribution in a confining potential}

\author{Matheus Girotto}
\email{matheus.girotto@ufrgs.br}
\affiliation{Instituto de F\'isica, Universidade Federal do Rio Grande do Sul, Caixa Postal 15051, CEP 91501-970, Porto Alegre, RS, Brazil}

\author{Alexandre P. dos Santos}
\email{alexandre.pereira@ufrgs.br}
\affiliation{Departamento de F\'isica, Universidade Federal de Santa Catarina, 88040-900, Florian\'opolis, Santa Catarina, Brazil}

\author{Yan Levin}
\email{levin@if.ufrgs.br}
\affiliation{Instituto de F\'isica, Universidade Federal do Rio Grande do Sul, Caixa Postal 15051, CEP 91501-970, Porto Alegre, RS, Brazil}

\begin{abstract}
We study a model of interacting vortices in a type II superconductor. 
In the weak coupling limit, we constructed a mean-field theory which allows
us to accurately calculate the vortex density distribution inside a confining potential. 
In the strong coupling limit, the correlations between the particles become important and the mean-field theory fails.  Contrary to recent suggestions, this does not imply failure of the 
Boltzmann-Gibbs statistical mechanics, as we clearly
demonstrate by comparing the results of Molecular Dynamics and Monte Carlo simulations. 

\end{abstract}

\maketitle

\section{Introduction}

Superconductivity is one of the greatest discoveries of the previous century. 
The practical applications of this phenomenon rely on
understanding the behavior of high temperatures superconductors~\cite{HeRo91,BlFe94,Br95,NaNa01} in a
magnetic field. Depending on the superconducting material and on external conditions, a phase with quantum magnetic vortices can appear. Superconductors with this property are called type II and are a subject of intense theoretical investigation~\cite{JeBr88,PlNo91,CoCl91,BrDo93a,BrDo93b,RiPl94,Co96,Ba97,ZaMo01,ZhDa04,LiLi09}. The Ginzburg-Landau theory~\cite{de89} predicts that the vortex-vortex interaction in a superconducting film has the form
\begin{equation}\label{V}
V(r)=q G({\bf x}_1,{\bf x}_2) \ ,
\end{equation}
where
\begin{equation}\label{green}
G({\bf x}_1,{\bf x}_2)=q K_0(\dfrac{|{\bf x}_1-{\bf x}_2|}{\lambda}) \ ,
\end{equation}
and $K_0$ is a modified Bessel function, $r=|{\bf x}_1-{\bf x}_2|$ is the distance between vortex~$1$ and vortex~$2$, $q$ is the vortex strength, and $\lambda$ is the London penetration length.  

A number of recent papers have studied the equilibrium distribution of vortices confined by an external 
potential~\cite{AnDa10,AnDa11,RiNo12}.  
The authors of these papers have argued that the ground state of these systems 
corresponds to the maximum of the non-extensive Tsallis entropy.  Contrary to this
suggestion, in this paper 
we will present a simple
mean-field theory which, in the framework of the usual Boltzmann-Gibbs (BG) statistical mechanics,  accounts very well for the equilibrium distribution of vortices.
Furthermore, comparing the results of Molecular Dynamics (MD) and Monte Carlo (MC) simulations,
we will show that the system of confined vortices is described by the standard BG 
statistical mechanics for all the coupling strengths.    

\section{Mean-Field Theory}

We will study a system of interacting vortices confined by an external potential
\begin{equation}\label{w}
W(x)=\alpha x^2/2 \ .
\end{equation} 
We first observe that 
the function, Eq.~\eqref{green}, satisfies a modified Helmholtz equation
\begin{equation}\label{gfe}
\nabla^2G({\bf x},{\bf x}_1)-G({\bf x},{\bf x}_1)=-2\pi q \delta({\bf x}-{\bf x}_1) \ ,
\end{equation}
where  all the lengths are now measured in units of $\lambda$.
Consider an infinite bi-dimensional system of vortices  in the $x$-$y$ plane, with periodic boundary conditions in the $y$ direction. The solution of Eq.~\eqref{gfe} can be expressed as~\cite{LePa11}
\begin{equation}\label{gf}
G({\bf x},{\bf x}_1)=\frac{\pi q}{L_y}\sum_{m=-\infty}^{\infty}e^{(\frac{2\pi mi}{L_y})(y-y_1)}\frac{e^{-\gamma_m|x-x_1|}}{\gamma_m} \ ,
\end{equation}
where
\begin{equation}
\gamma_m=\sqrt{1+4\pi^2m^2/L_y^2} \ ,
\end{equation}
$m$ are integers and $L_y$ is the width of the periodic stripe in the $y$ direction.

In equilibrium, the particle distribution is given by 
\begin{equation}\label{om}
\rho(x)=A  e^{-\beta \omega (x)} \ ,
\end{equation}
where $\beta=1/k_B T$, $\omega (x)$ is the potential of mean force (PMF), and $A$ is the normalization constant~\cite{Le02}.  

\begin{figure}[t]
\begin{center}
\includegraphics[width=8.cm]{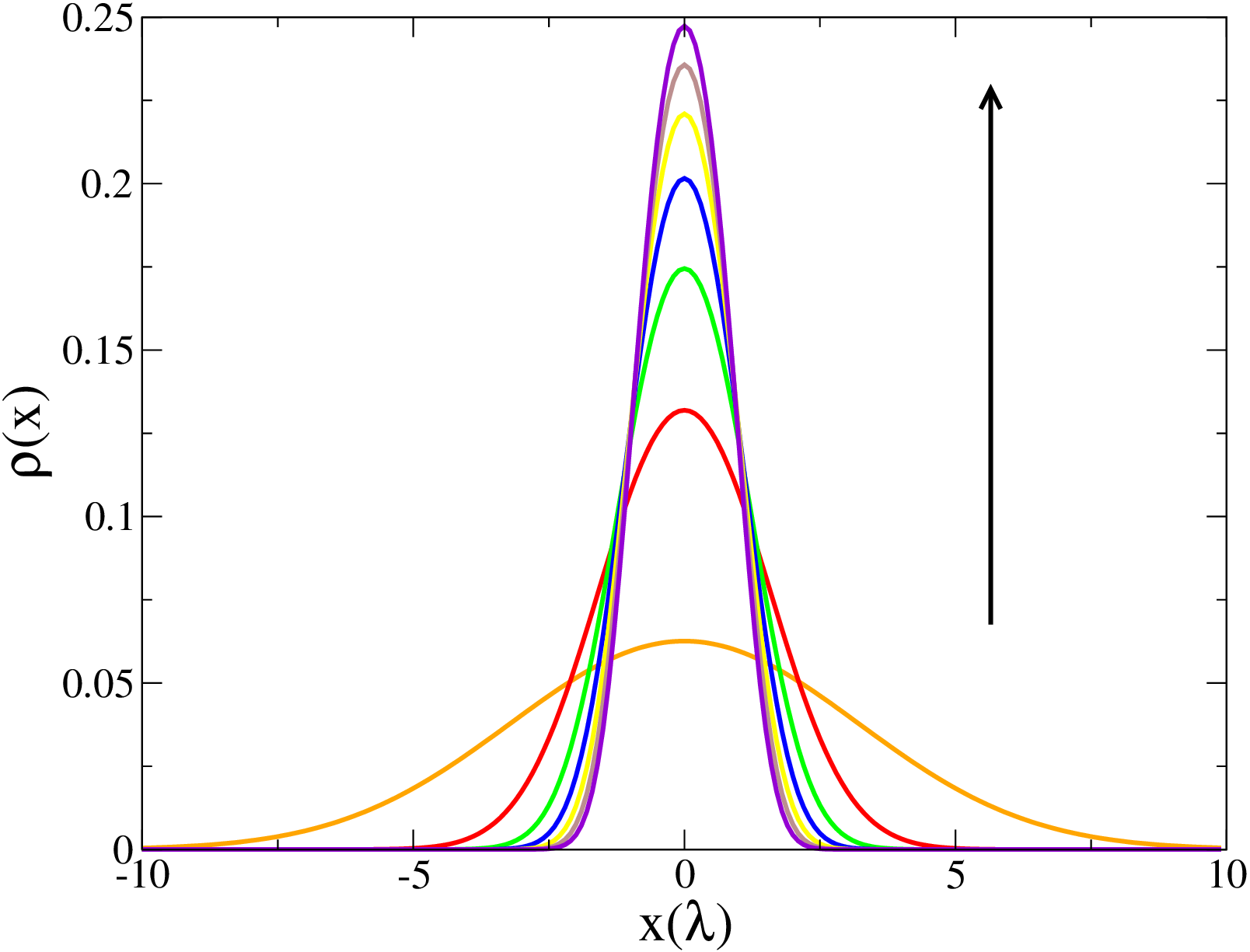}
\end{center}
\caption{Theoretical density profiles for confined vortices. The arrow indicates increasing coupling strength (lowering temperature): $\epsilon=\chi=0.1$, $0.5$, $1$, $1.5$, $2$, $2.5$ and $3$.}
\label{fig3}
\end{figure}

In the weak-coupling limit (high temperatures), the correlations between the vortices can  
be neglected and the PMF can be approximated by  $\omega(x)=q\phi(x)+W(x)$.  
The particle distribution then becomes
\begin{equation}\label{dens}
\rho(x)=A  e^{-\beta[q\phi(x)+W(x)]} \ ,
\end{equation}
where $A$ is 
\begin{equation}
A=\frac{N}{L_y\int_{-\infty}^{+\infty} dx \ e^{-\beta[q\phi(x)+W(x)]}} \ .
\end{equation}
The potential $\phi(x)$ can be calculated using the Green's function 
\begin{equation}\label{phi1}
\phi( x)=\int d{\bf x}' \rho( x')G({\bf x},{\bf x}') \ .
\end{equation}
Eq.~\eqref{gf} allows us to rewrite Eq.~\eqref{phi1} as
\begin{equation}\label{phi2}
\phi(x)=\pi q\int_{-\infty}^{+\infty} dx' \rho(x')e^{-|x'-x|} \ .
\end{equation}
The two equations  \eqref{dens} and \eqref{phi2} can be solved iteratively.
To quantify the strength of the vortex-vortex  and the trap-vortex interaction, 
it is convenient to define the following dimensionless parameters
\begin{equation}
\epsilon=\frac{q^2}{k_BT} \quad\mbox{and}\quad \chi=\frac{\alpha \lambda^2}{k_BT} \ .
\end{equation}
In Fig.~\ref{fig3} we present solutions of Eqs.\eqref{dens} and \eqref{phi2} 
for various coupling parameters. 

\section{Molecular Dynamics Simulations}

To verify the predictions of the mean-field theory, we first perform  MD simulations. 
A system of $N=200$ vortices interacting  by the pair potential, Eq.~\eqref{V},
is confined inside an infinite stripe of width $L_y=400$,  
with a trap potential $W(x)$ acting along the $x$ direction.
As in the references \cite{AnDa10,AnDa11}, periodic boundary conditions are used in the $y$ direction.
The equations of motion for each particle $i$,
\begin{equation}
d^2 {\bf x}_i/dt^2=-\sum_{j=1}^{N}\nabla_{{\bf x}_i} q G({\bf x}_i,{\bf x}_j) -\alpha x_i\hat{\bf e}_x \ ,
\end{equation}
are integrated using the leapfrog algorithm.

\begin{figure}[t]
\includegraphics[width=5.7cm]{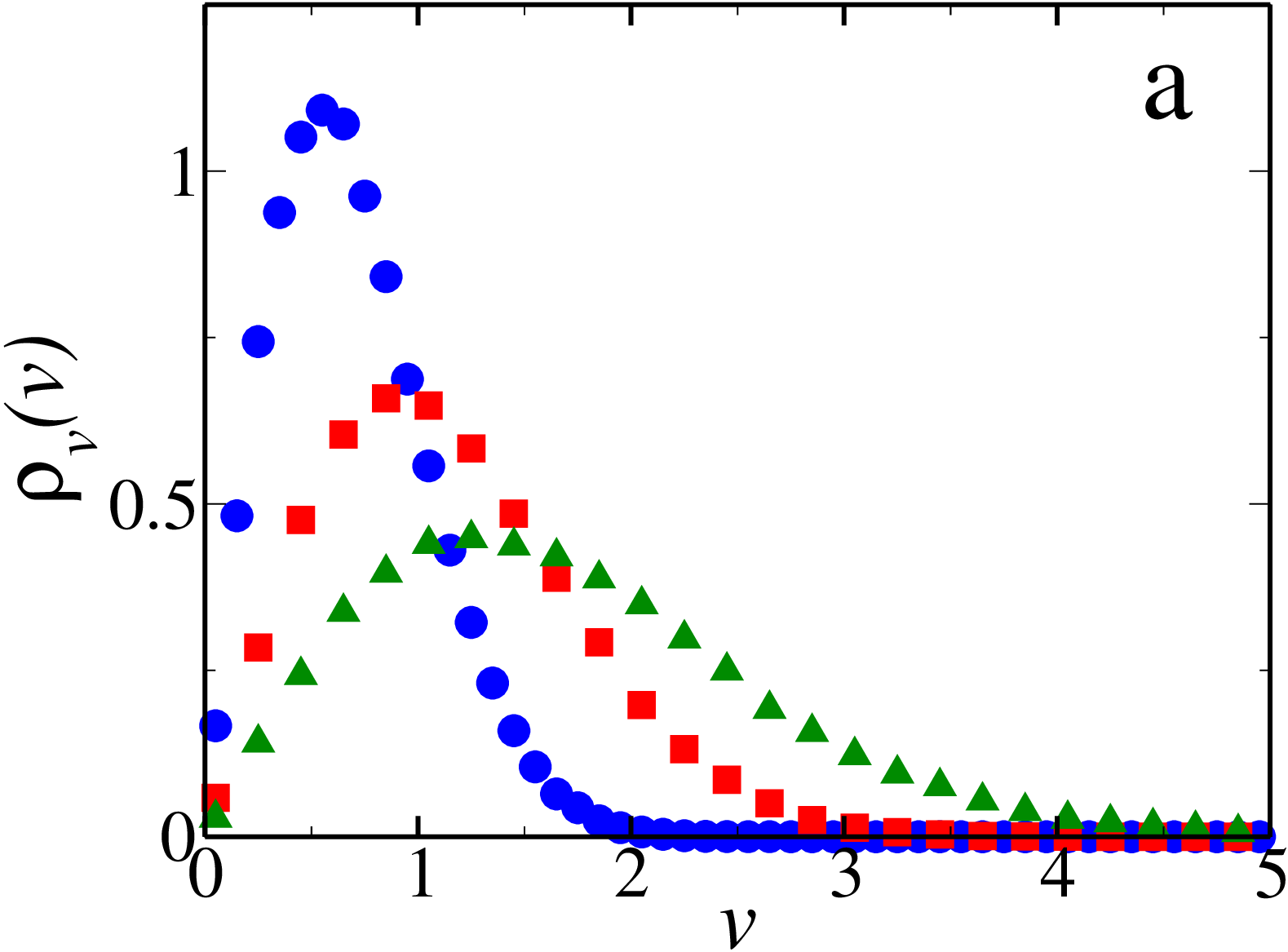}\vspace{0.2cm}
\includegraphics[width=5.7cm]{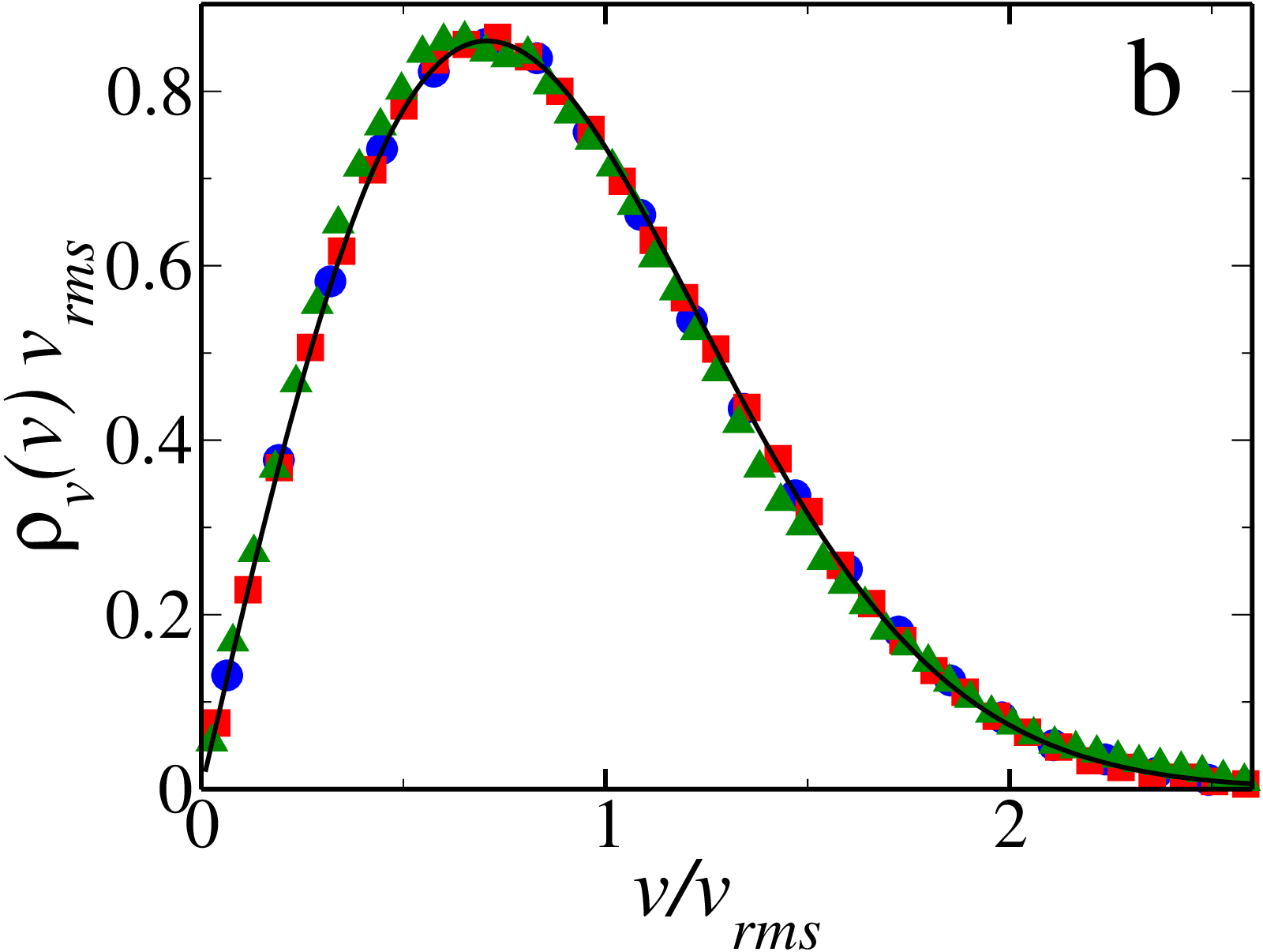}\vspace{0.2cm}
\caption{(a) Velocity distributions for various initial conditions:  circles, $\epsilon=3.26$ and $\chi=1.63$; squares, $\epsilon=1.2$ and $\chi=0.6$; triangles, $\epsilon=0.54$ and $\chi=0.27$. 
Panel (b) shows that when the velocity is scaled with $v_{rms}$ and
the distribution function is scaled with $1/v_{rms}$, all the curves collapse onto the universal 2D
Maxwell-Boltzmann distribution $f(x)=2 x e^{-x^2}$, represented by the solid line.}
\label{fig1}
\end{figure}

In the simulations, a system is prepared in various initial conditions and is allowed to relax 
until a stationary particle distribution is established. After the equilibrium is achieved, 
we calculate the distribution of particle velocities, shown in Fig.~\ref{fig1}a. If the stationary state is the usual
BG equilibrium, we expect the particle distribution to have the Maxwell-Boltzmann form, which in 2D is 
\begin{equation}\label{mbd}
\rho_{v}(v)= \dfrac{2v}{v_{rms}^2} e^{-v^2/v_{rms}^2} \ ,
\end{equation}
with $v_{rms}=\sqrt{\langle v^2 \rangle}$.  This means that if the velocities are
scaled with $v_{rms}$ and the distribution is scaled with $1/v_{rms}$, all the curves plotted
in Fig \ref{fig1}a should collapse onto one universal curve $f(x)=2 x e^{-x^2}$.  This is precisely
what is shown to happen in Fig.~\ref{fig1}b.
To obtain the density distribution using MD simulation, we divide the simulation stripe into bins of width $\Delta_x$, and calculate the average number of particles in each bin.

\begin{figure}[h]
\includegraphics[width=5.7cm]{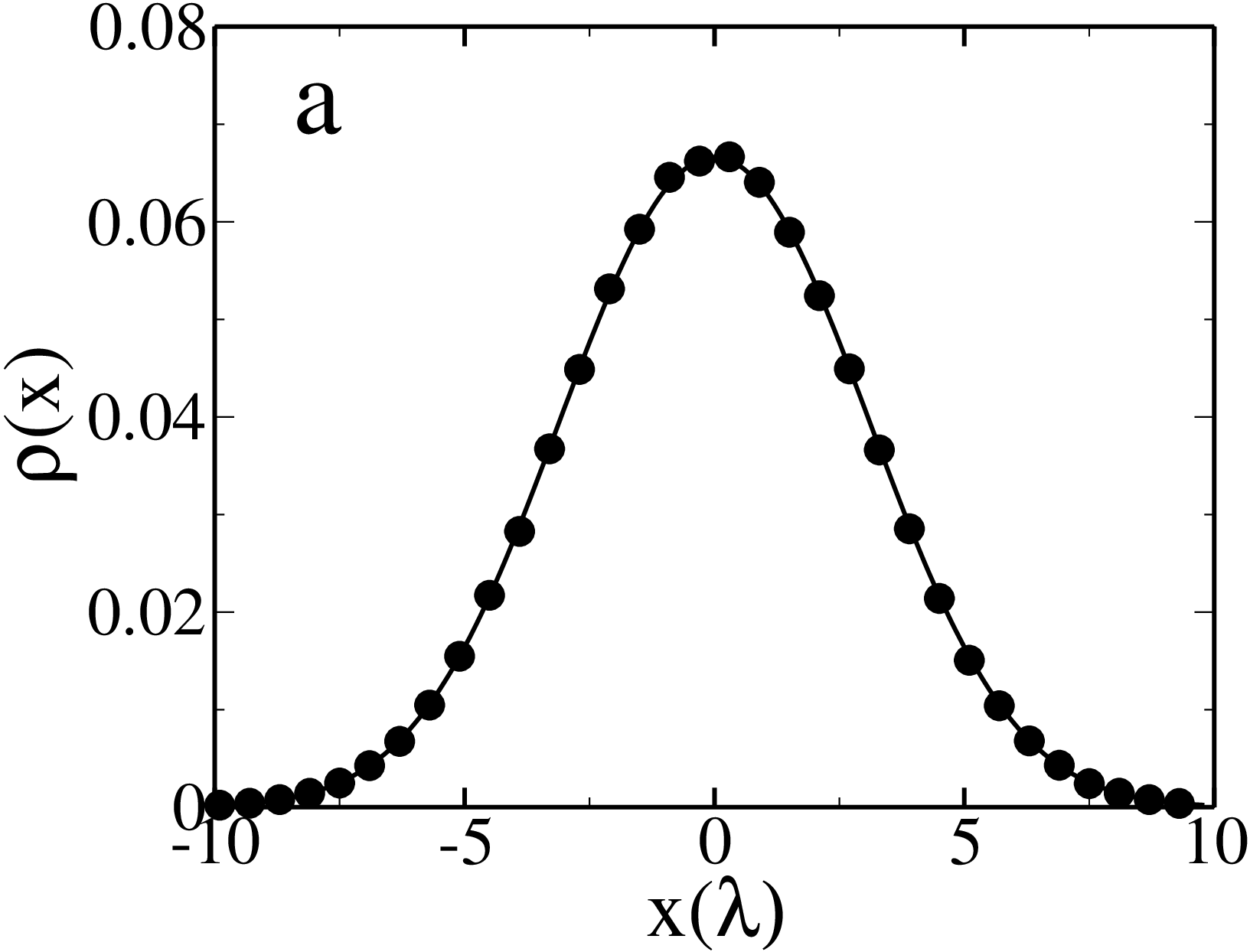}\vspace{0.2cm}
\includegraphics[width=5.7cm]{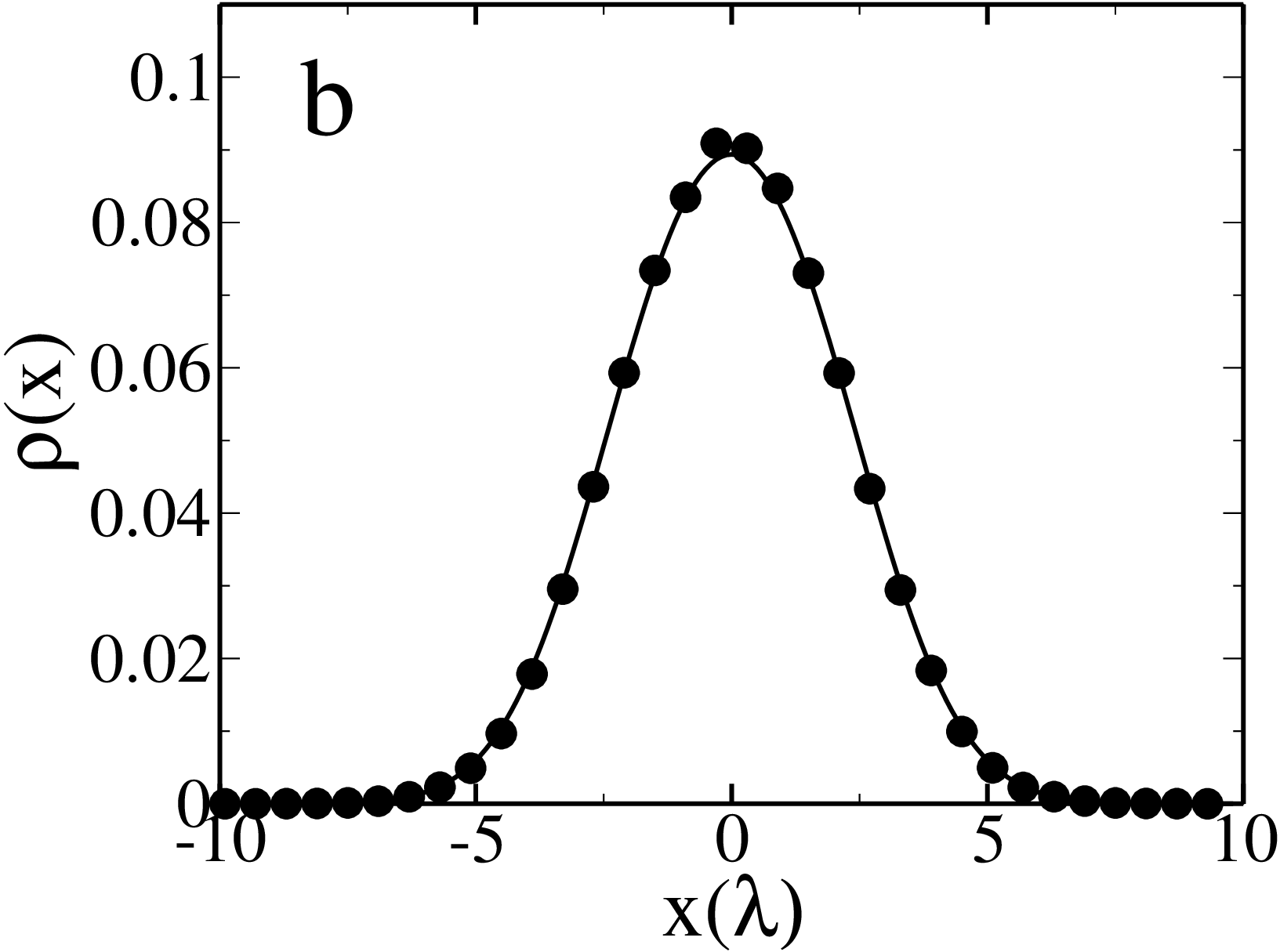}\vspace{0.2cm}
\includegraphics[width=5.7cm]{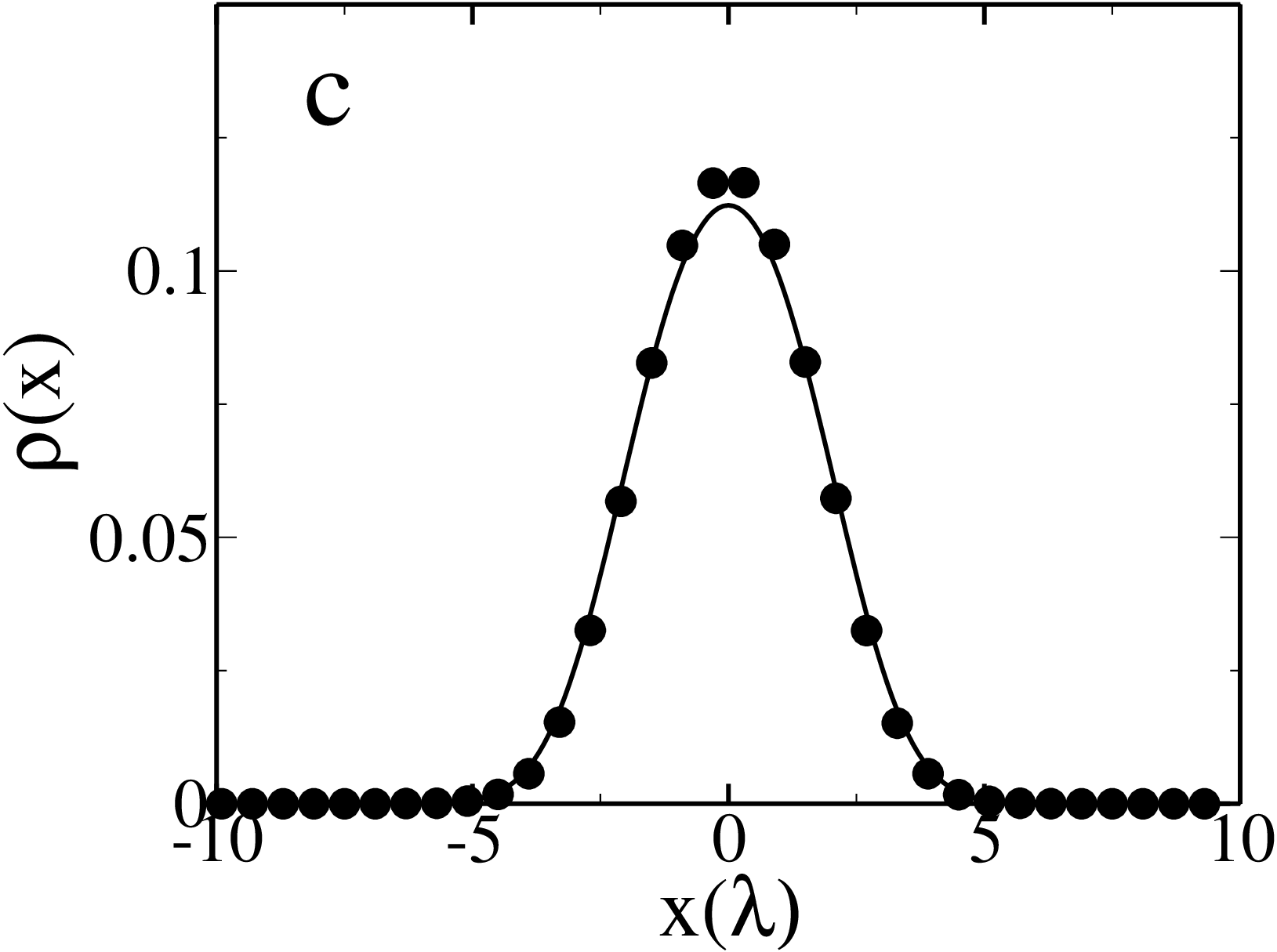}\vspace{0.2cm}
\includegraphics[width=5.7cm]{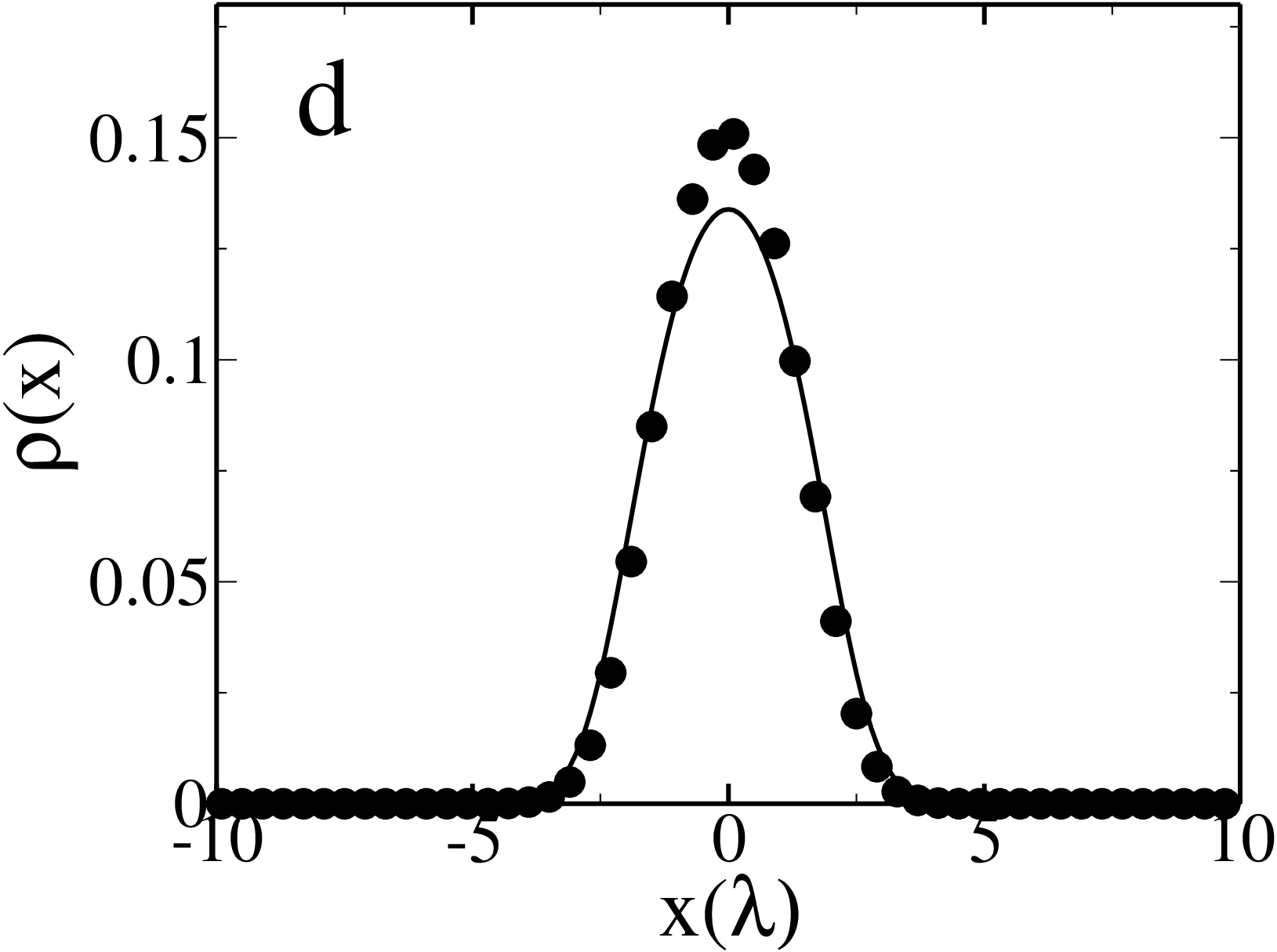}
\caption{Vortex density profiles, symbols represent the MD simulation data and lines the 
predictions of the mean-field theory. For lower  
temperatures, the parameters $\epsilon$ and $\chi$ increase. The panels correspond to parameters: (a) $\epsilon=0.49$ and $\chi=0.1225$; (b) $\epsilon=1$ and $\chi=0.25$; (c) $\epsilon=1.96$ and $\chi=0.49$; and (d) $\epsilon=4$ and $\chi=1$.}
\label{fig4}
\end{figure}

\begin{figure}[t]
\begin{center}
\includegraphics[width=8.cm]{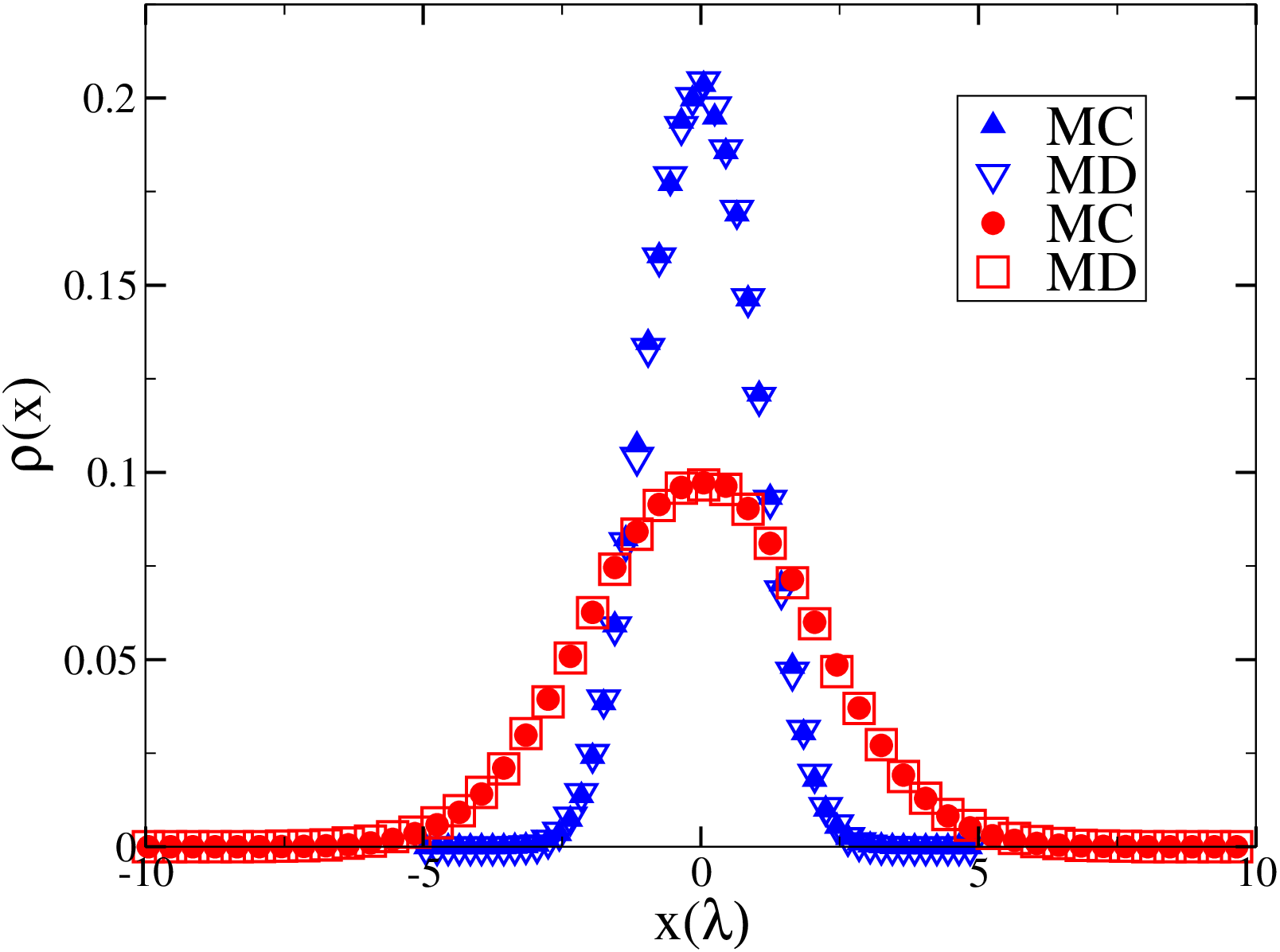}
\end{center}
\caption{Density profiles for confined vortices obtained using MD and MC simulations:  up and down triangles, $\epsilon=0.54$ and $\chi=0.27$; circles and squares, $\epsilon=3.26$ and $\chi=1.63$.  
MC and MD predict identical vortex density distributions.}
\label{fig2}
\end{figure}

To compare the predictions of the mean-field theory with the results of MD simulations,
we let the system relax to equilibrium and calculate the $\langle v^2 \rangle$.
For  systems with short range interactions the canonical and the micro-canonical ensembles must be equivalent, so that in 2D, $m v_{rms}^2/2=k_B T$.   
Using this temperature, the mean-field vortex distribution can be calculated using Eq. \eqref{dens}.
Comparing the predictions of the mean-field theory with the results of MD simulations, we see that for high temperatures there is an excellent agreement, see Fig.~\ref{fig4}.  In this limit the mean-field theory, Eq. \eqref{dens}, becomes exact~\cite{Le02}. On the other hand, in the strong coupling limit (low temperatures), the correlations between the particles are important and significant deviations from the results of simulations can be seen.  Correlations lead to a larger concentration of particles in the low energy states than is predicted by the mean-field theory~\cite{Le02}.  This is similar to the process of overcharging observed in colloidal suspensions with multivalent ions~\cite{GrNg02,PiBa05,DoDi10a}.

Andrade~\textit{et al.}~\cite{AnDa10,AnDa11} and Ribeiro~\textit{et al.}~\cite{RiNo12} have argued that at low temperatures, the vortices in a type II superconductor obey Tsallis statistics~(TS).  In particular, they claimed that the ground state of interacting vortices in a confining potential corresponds to the maximum of the Tsallis entropy.  The arguments of Andrade~\textit{et al.} are based on a solution of an {\it approximate}  Fokker-Planck equation. This equation is very interesting 
and allows to make some important predictions, such as 
front propagation in type II superconductors~\cite{ZaMo01}.  However, the fact that the stationary solution of this {\it approximate}  
equation at $T=0$ is a "q-Gaussian" does not provide any justification for the relevance of the 
non-extensive statistical mechanics to thermodynamics of superconducting
vortices.  In fact, the Fokker-Planck equation for vortex density
is an approximation of a more accurate Nernst-Planck-like equation, which does have the usual Boltzmann distribution
as a stationary state.  Neither of these equations, however, take into account the correlations between the particles, so that
both can only be valid in the mean-field limit.  Nevertheless, even in this
limit, Ref. \cite{LePa11} shows that the q-Gaussian solution of the Fokker-Planck equation obtained by Andrade~\textit{et al.} is inconsistent with the solution of the more 
accurate Nernst-Planck equation.

To see that the equilibrium state of the system studied by Andrade~\textit{et al.} 
is indeed described by the usual BG statistical mechanics for any temperature, we perform a series of MD and MC simulations.  In MC simulations, we use the usual Metropolis algorithm~\cite{MeRo53} which is constructed to evolve the system through a Markov process towards a stationary state in which the particles are distributed (in the phase space) according to the Boltzmann distribution.  Clearly if the agreement between  MD and MC simulations is found, it will unequivocally show that the system of vortices interacting by the potential of Eq.~\eqref{V}, is both ergodic and mixing and is 
described by the usual BG statistical mechanics.  

\begin{figure}
\includegraphics[width=5.7cm]{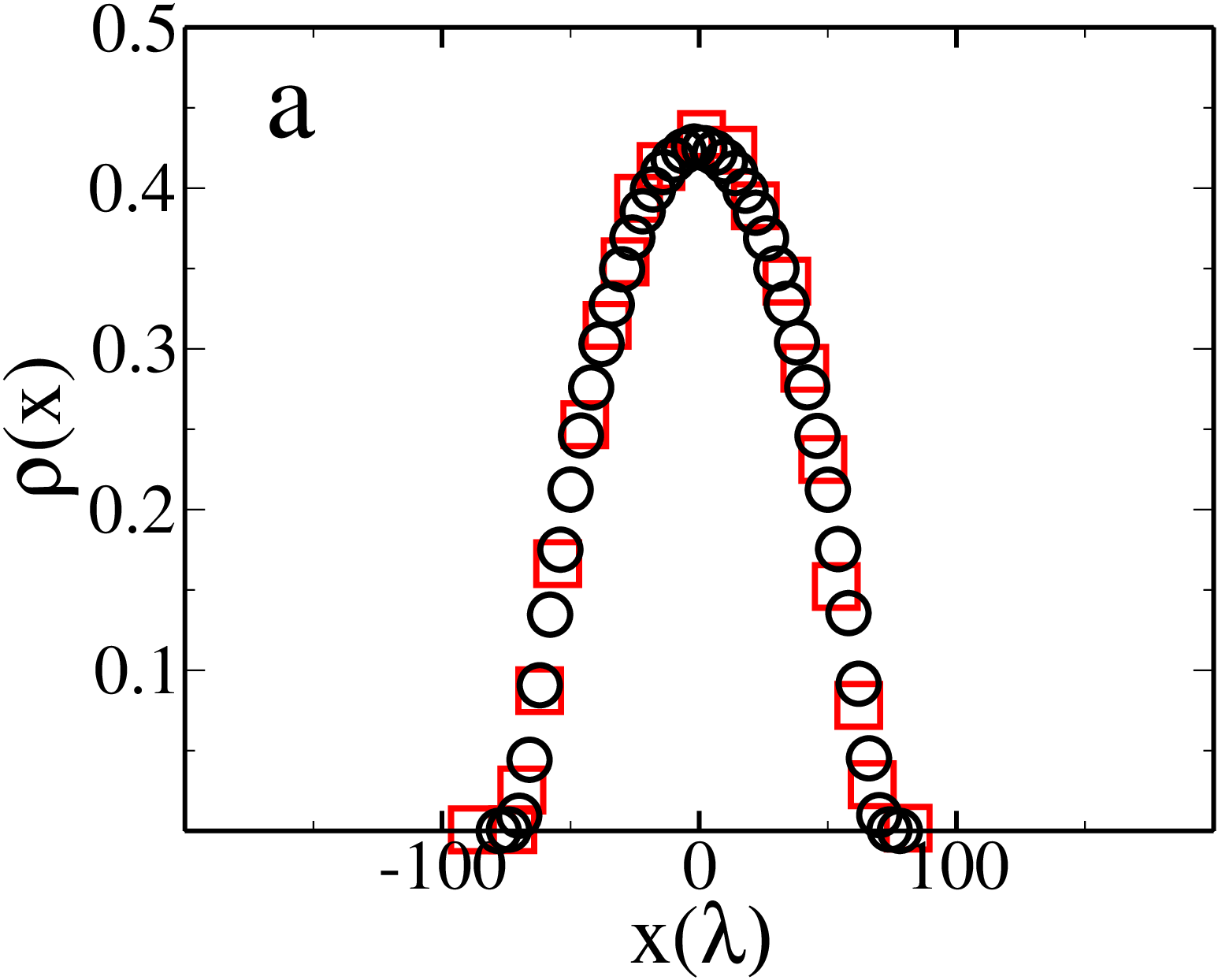}\vspace{0.2cm}
\includegraphics[width=5.7cm]{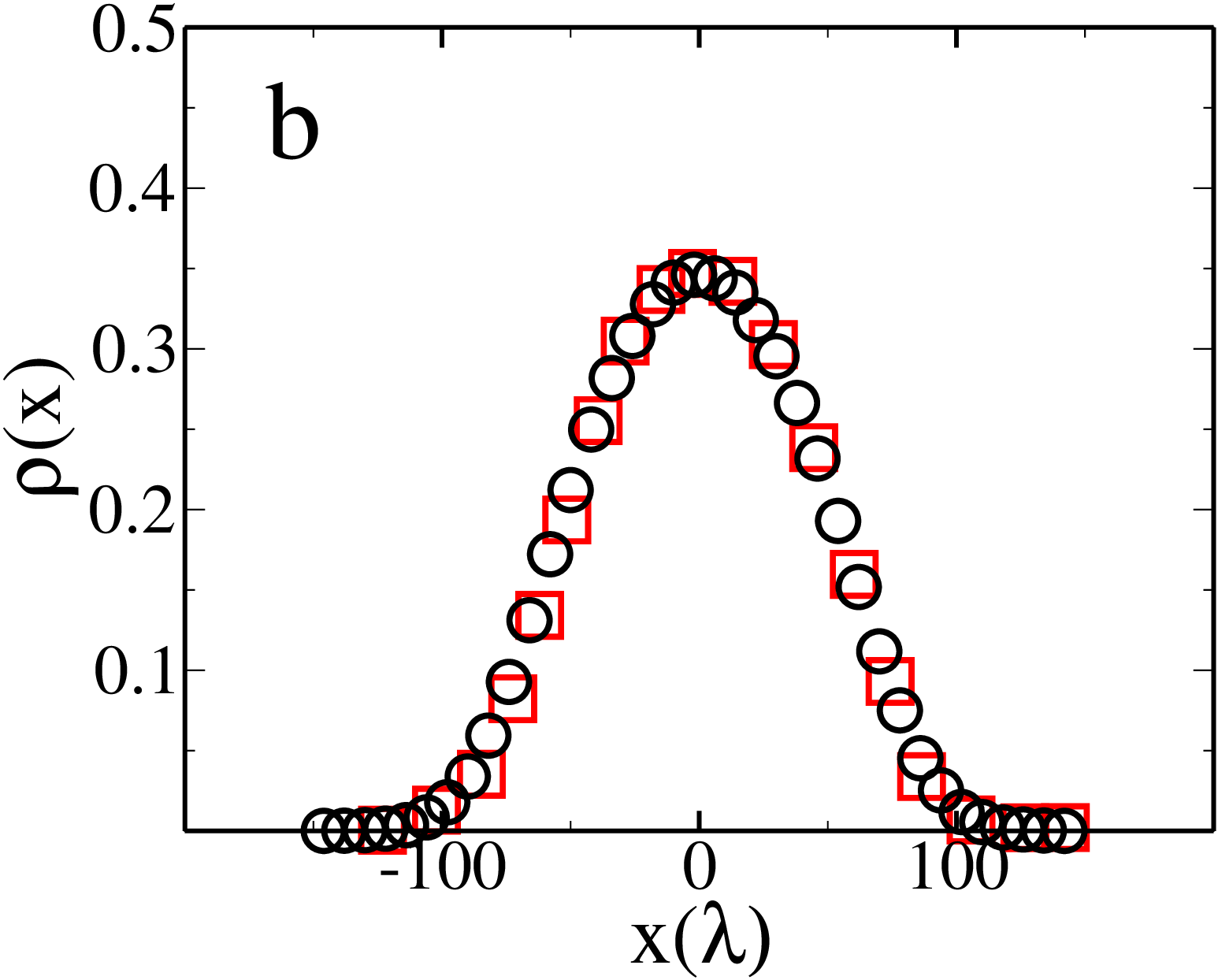}\vspace{0.2cm}
\includegraphics[width=5.7cm]{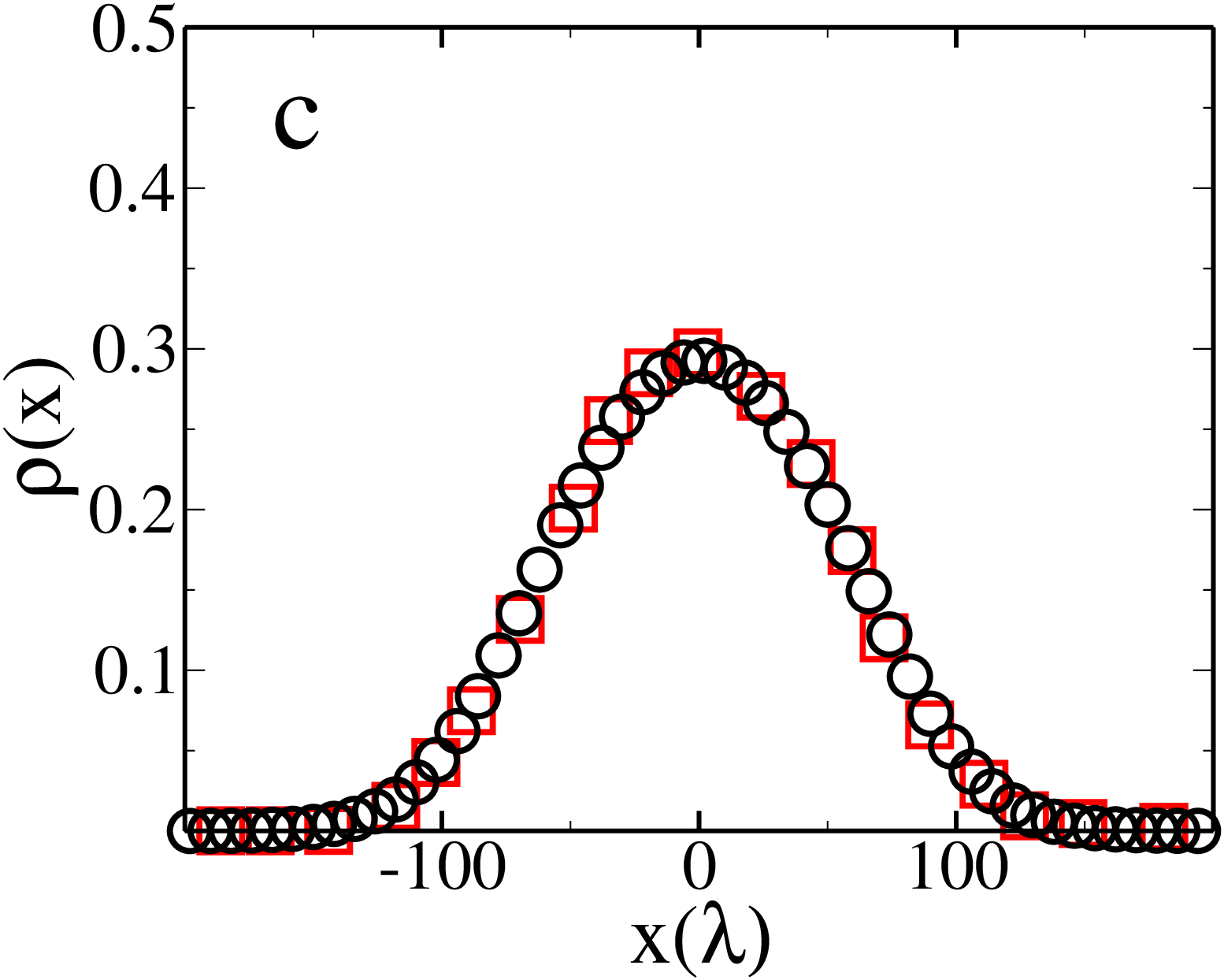}\vspace{0.2cm}
\includegraphics[width=5.7cm]{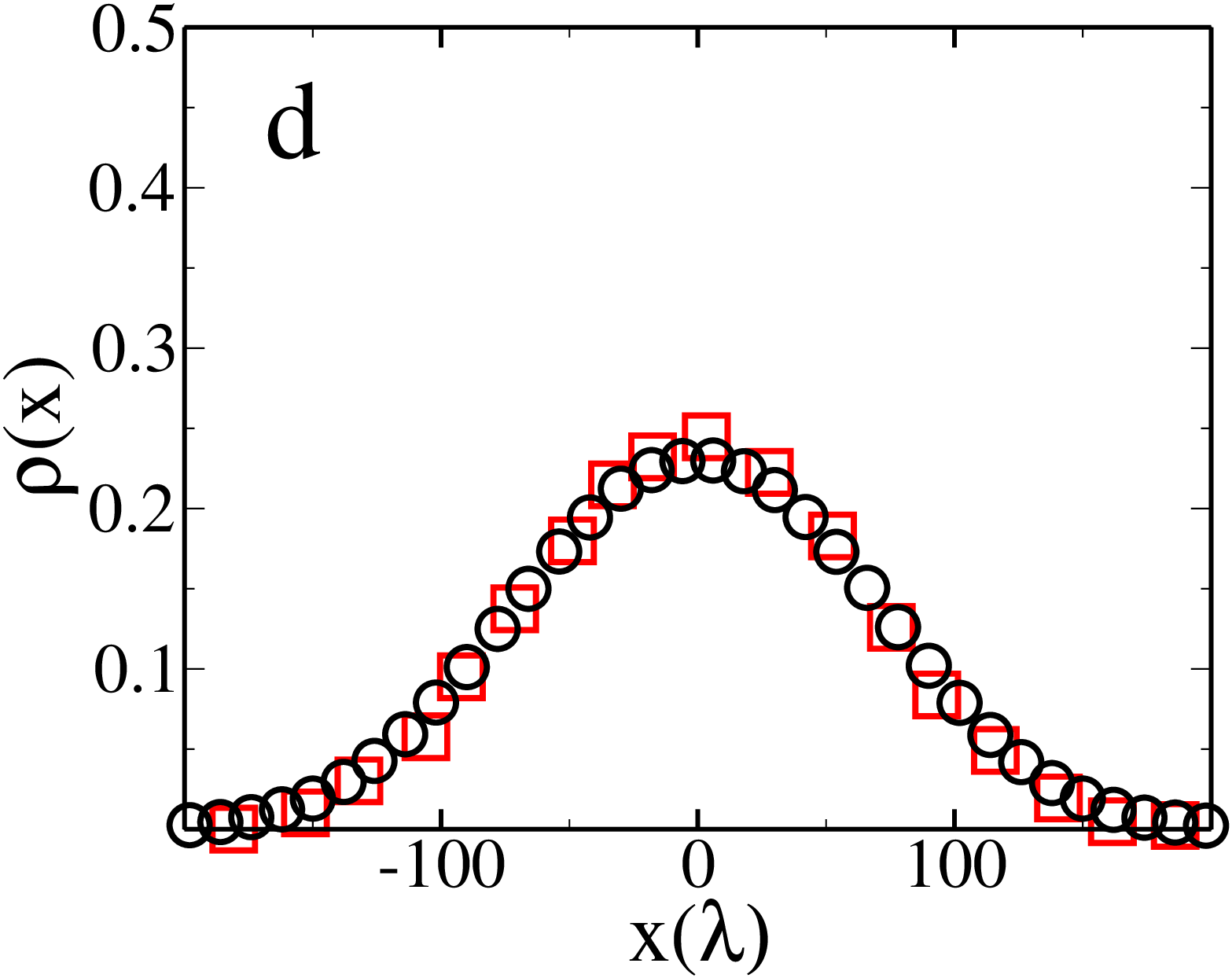}
\caption{Vortex density profiles. Squares are the data of Andrade~\textit{et al.} from Fig.2 of Ref.~\cite{AnDa10} and circles are the results of our MC simulations. The parameters are: $\alpha=10^{-3}~q^2/\lambda^2$, $N=800$, $L_y=20 \lambda$: (a)~$T=0.1~q^2/k_B$, (b)~$T=1.0~q^2/k_B$, (c)~$T=2.0~q^2/k_B$ and (d)~$T=4.0~q^2/k_B$.  The agreement between the two simulations clearly shows
that the stationary state to which the system relaxes is the usual BG equilibrium.}
\label{fig5}
\end{figure}
\section{Monte Carlo simulations}

We have seen already that the vortex velocity distribution is in perfect
agreement with 
the BG statistical mechanics.  In this section, we will show that the vortex density distribution
is also described by the BG statistical mechanics. To do this we perform MC simulations and compare them with the results of MD simulations. MC simulations are designed to force the particles into an equilibrium state corresponding to the maximum of the Boltzmann entropy (in the microcanonical ensemble) or the minimum of the Helmholtz free energy, in the canonical ensemble.  
To simulate canonical ensemble one can use the Metropolis algorithm.
In the Markov chain of the Metropolis algorithm, a new  configuration $n$ is constructed  from an old configuration $o$ by a small displacement of a random particle. The new state is accepted with a probability $P=\min\{1,e^{-\beta(E_n-E_o)}\}$, where $\beta=1/k_B T$. 
If the movement is not accepted, the configuration $o$ is preserved and counted as a new state. 
The length of the displacement is adjusted during the simulation in order to obtain the acceptance rate of $50\%$. 
The energy of the system used in the MC simulations is given by
\begin{equation}
E=\sum_{i=1}^{N-1}\sum_{j=i+1}^{N}q G({\bf x}_i,{\bf x}_j)+\sum_{i=1}^{N}W(x_i) \ .
\end{equation}
Metropolis algorithm insures that the system evolves to the BG thermodynamic equilibrium. 
The averages are calculated using $10^5$ uncorrelated states, obtained after $10^6$ MC steps for equilibration. 
Fig.~\ref{fig2} shows a perfect agreement between the results of our microcanonical MD and canonical MC simulations.  
In Fig. \ref{fig5} we compare the results of
our MC simulations with the simulations of  Andrade~\textit{et al.} (Fig.2 of Ref.~\cite{AnDa10}) performed 
using an overdamped dynamics with a thermostat.  Once again, the two are indistinguishable.  
This unequivocally demonstrates that the system of vortices, interacting by the potential of Eq.~\eqref{V} is  
described by the usual BG statistical mechanics.

\section{Conclusions}

We have studied a simple model of interacting vortices in a type II superconductor. 
In the weak coupling limit we have constructed a mean-field theory which allows
us to accurately calculate the equilibrium vortex density distribution inside a confining potential. 
In the strong coupling limit the correlations between the particles become important and the mean-field theory fails.  This, however, does not imply the failure of the BG statistics, as is clearly demonstrated by the perfect agreement between MD and MC simulations and by the Maxwell-Boltzmann distribution of the particle velocities. 

It is very difficult to study theoretically the correlations in inhomogeneous liquids. A number of different approaches, such as density functional  theory (DFT)~\cite{Ta85,CuAs85,DeAs89,Gr91,DiTa99} and integral equations~\cite{Kj88a,Kj88b},  have been developed over the years. All these theoretical methods are firmly embedded in the framework of the BG statistics. Introduction of "novel type" of entropies~\cite{AnDa10,AnDa11,RiNo12}  as a way to ``fit in" the inter-particle correlations does not help to shed any new light on the equilibrium properties of these interesting systems.

\bibliography{paper.bbl}

\end{document}